\documentclass[utf8]{frontiersSCNS} 

\usepackage{url,hyperref,lineno,microtype,subcaption}
\usepackage[onehalfspacing]{setspace}
\usepackage{multicol}
\usepackage[final]{changes}
\definechangesauthor[name=BC, color=blue]{BC}
\definechangesauthor[name=OW, color=blue]{OW}
\definechangesauthor[name=OW, color=blue]{CJ}
\definechangesauthor[name=R1, color=red]{R1}
\definechangesauthor[name=R2, color=red]{R2}
\definechangesauthor[name=A1, color=blue]{A1}
\definechangesauthor[name=A2, color=blue]{A2}

\def\keyFont{\fontsize{8}{11}\helveticabold }
\def\firstAuthorLast{Cecconi {et~al.}} 
\def\Authors{B. Cecconi\,$^{1,*}$, O. Witasse\,$^{2}$, C.M. Jackman\,$^{3}$, 
B. S\'anchez-Cano\,$^{4}$, and M. L. Mays\,$^{5}$}
\def\Address{$^{1}$ LESIA, Observatoire de Paris, Université PSL, 
Sorbonne Université, Université de Paris, CNRS, Meudon, France\\
$^{2}$ ESTEC-Scientific Support Office, European Space Agency, Noordwijk, 
Netherlands\\
$^{3}$ School of Cosmic Physics, DIAS, Dublin, Ireland\\
$^{4}$ School of Physics and Astronomy, University of Leicester, Leicester, 
UK\\
$^{5}$ CCMC, NASA/GSFC, Code 674, Greenbelt, MD 20771, USA}
\def\corrAuthor{B. Cecconi}

\def\corrEmail{baptiste.cecconi@obspm.fr}

\begin{document}
\onecolumn
\firstpage{1}

\title[Effect of an ICME on SKR]{Effect of an interplanetary coronal mass ejection 
on Saturn's radio emission} 

\author[\firstAuthorLast ]{\Authors} 
\address{} 
\correspondance{} 

\extraAuth{}

\maketitle

\begin{abstract}

The Saturn  Kilometric  Radiation (SKR)  was  observed  for  the  first  time  
during the flyby of Saturn by the Voyager spacecraft in 1980. These radio 
emissions, in the range of a few kHz to 1 MHz, are emitted by electrons 
travelling around auroral magnetic field lines. Their study is useful to 
understand the variability of a magnetosphere and its coupling with the solar
wind. Previous studies have shown a strong correlation between the solar wind
\added[id=R1]{dynamic} pressure and the SKR intensity. However, up to now, the effect of an
Interplanetary Coronal Mass Ejection (ICME) has never been examined in detail, 
due to the lack of SKR observations at the time when an ICME can be tracked and
its different parts be clearly identified. In this study, we take advantage of a
large ICME that reached Saturn mid-November 2014 \citep{Witasse:2017jv}.
At that time, the Cassini spacecraft was fortunately travelling within the solar
wind for a few days, and provided a very accurate timing of the ICME structure. A
survey of the Cassini data for the same period indicated a significant increase in
the SKR emissions, showing a good correlation after the passage of the ICME shock
with a delay of $\sim$13 hours and after the magnetic cloud passage with a delay of
25-42 hours. \deleted[id=BC]{In between, a smaller SKR burst could be correlated with a proton
flux peak occurring during the passage of the ejecta of the ICME.}

\tiny
 \keyFont{ \section{Keywords:} Saturn, Cassini, SKR emission, Interplanetary 
 Coronal Mass Ejection, solar wind} 
\end{abstract}

\section{Introduction}

The Saturn Kilometric Radiation (SKR) was observed for the first time by 
\cite{kaiser_Sci_80} during the flyby of Saturn in January 1980, with the 
Voyager/PRA (Planetary Radio Astronomy) instrument \citep{warwick_SSR_77}. The 
SKR radio emission is a low frequency non-thermal radio emission observed between 
a few kHz to about 1 MHz. It is emitted through the Cyclotron Maser Instability \added[id=BC]{(CMI)}
\citep{wulee_ApJ_79, lamy_GRL_10b} triggered by accelerated electrons travelling 
around auroral field lines \citep{lamy_JGR_09}. SKR is thus directly related to
magnetospheric dynamical activity.

The analysis of Voyager PRA data showed a strong correlation between the 
\replaced[id=R1]{solar wind}{Solar Wind} \added[id=R1]{dynamic} pressure and SKR emitted flux 
\citep{Desch:1982ta,desch_JGR_83}. \citet{Desch:2007bx} even showed 
\added[id=R1]{that} the SKR disappeared while Saturn passed through the tail of 
the Jovian magnetosphere. Thanks to the Cassini RPWS (Radio and Plasma 
Waves Science) experiment \citep{gurnett_SSR_04} we have quasi-continuous 
observations of the SKR from early 2004 to the end of mission in October 
2017. The radio data have the potential to be used as a remote proxy for 
both upstream driving and magnetospheric dynamics, and many studies have 
sought to explore the role of the solar wind perturbations in stimulating 
auroral and radio phenomena. Corotating interaction region (CIR) compressions 
in the solar wind have been found to intensify the SKR bursts and occasionally 
to disrupt their regular phasing \citep{badman_AG_08,Kurth:2016iu}. Dynamic 
magnetotail reconnection events have also been shown to correlate well with
intensifications and low frequency extensions (LFEs) of the SKR 
\citep{bunce05,jackman_JGR_05,jackman_JGR_09,Reed18}.

\begin{figure}[h]
    \centering\includegraphics[width=0.8\linewidth]{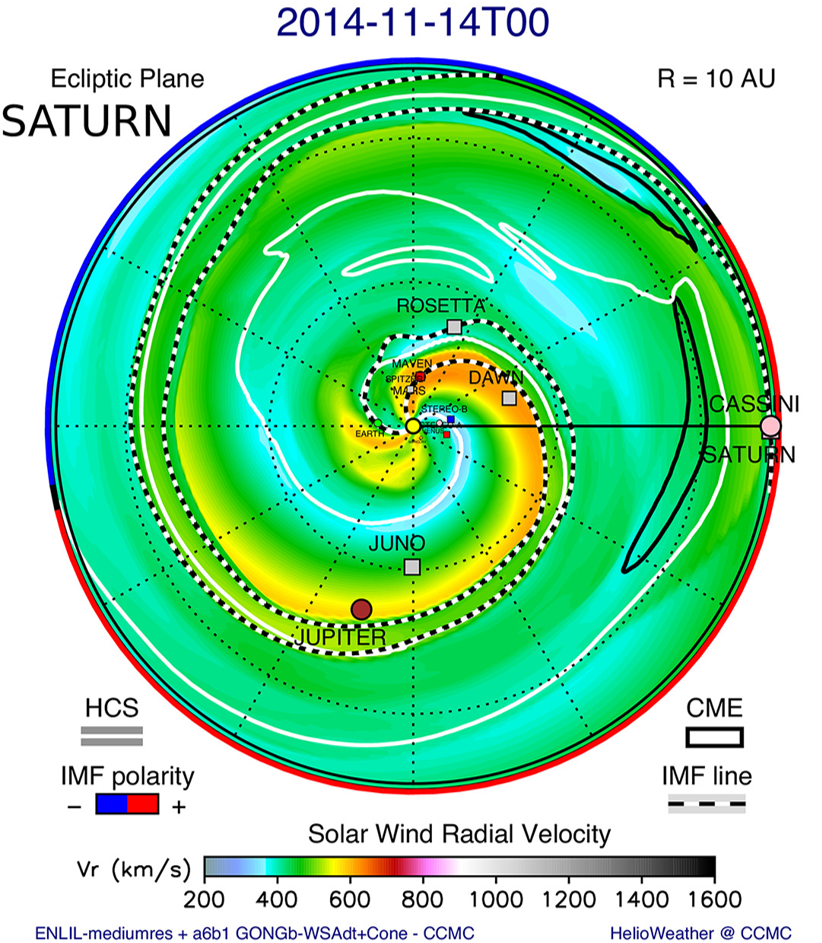}
    \caption{\replaced[id=R1]{Solar }{Still of the solar} wind velocity in the ecliptic 
    plane from a WSA‐ENLIL + Cone model simulation showing the ICME propagation 
    at Saturn. The ICME is marked as a black \replaced[id=R1]{contour}{bubble} on the right side 
    before Saturn. Time is given in the format YYYY-MM-DDThh. The full simulation 
    is available at \citep{Witasse:2017jv} and \url{http://ccmc.gsfc.nasa.gov} 
    under run ID \texttt{Leila\_Mays\_092716\_SH\_1}}
    \label{fig:enlil_still}
\end{figure}

\replaced[id=BC]{The }{Although the} link between enhanced solar wind 
\replaced[id=BC]{parameters }{dynamic pressure} and intense \replaced[id=BC]{planetary auroral activity }{SKR} 
has been known for many years \added[id=BC]{at Saturn} 
\replaced[id=BC]{\citep[e.g.,][]{kaiser_Sci_80,desch_JGR_83,Rucker:1984ws,Crary:2005eu,Taubenschuss06,
Lamy:2018be} }{\mbox{\citep[e.g.,][]{kaiser_Sci_80,desch_JGR_83,Crary:2005eu,Taubenschuss06}}},
\added[id=BC]{and has also been studied in the context of Jupiter \citep{prange_Nat_04,Hess:2012dx,Kita:2019ei} 
or Uranus \citep{Lamy:2012jb}, using modelled solar wind parameters when \emph{in-situ} observations were not
available}. \added[id=BC]{In the case of Saturn, the solar wind dynamic pressure has been identified as 
the main driver of enhanced SKR activity \citep{kaiser_Sci_80,Rucker:1984ws}.} \replaced[id=BC]{The }{the} 
effect of Interplanetary Coronal Mass Ejections (ICME) \added[id=BC]{(see, e.g., \citet{Zurbuchen:2006hm}
for detailed descriptions of ICME properties)} on SKR activity has never been
described in detail, due to the lack of SKR observations at the time when an ICME
impact can be clearly identified. In this study, we take advantage of a large ICME
that reached Saturn in November 2014 \citep{Witasse:2017jv}. The solar wind
feature was tracked from the Sun up to Voyager 2 as well as simulated with the
WSA-ENLIL+Cone model. A \replaced[id=R1]{snapshot }{still} of this model at Saturn is shown 
in Figure \ref{fig:enlil_still}. One of the major challenges for interpreting 
remote sensing observations is the lack of an upstream monitor to provide context 
for the observations and a quantitative measure of the solar wind driving conditions. 
However, for this rare example, the Cassini spacecraft was situated upstream of 
the bow shock, sampling the solar wind for a few days. Thus, it has been possible 
to use in situ data from the Cassini magnetometer (MAG) \citep{Dougherty:2004tr} 
and energetic particle instrument (MIMI) \citep{Krimigis:2004cf} to provide 
accurate timing for the ICME arrival. The ICME impacted Cassini on November 12th, 
2014. A survey of the RPWS data for the same period indicated a significant 
increase in the SKR emissions. This article reports on the correlation between 
this solar wind disturbance and SKR emissions.

\section{Data and trajectories}
\label{sec:data-traj}
In this study, we use several datasets from various instruments: radio frequency 
data from the High Frequency Receiver (HFR) of Cassini/RPWS 
\citep{gurnett_SSR_04}; magnetic field data from the Cassini/MAG (Magnetometer)
\citep{Dougherty:2004tr}; and particle \replaced[id=R1]{observations }{distribution function moments} from
Cassini/MIMI (Magnetosphere Imaging Instrument) \citep{Krimigis:2004cf}. The
trajectory data is provided by the CDPP (Centre de Données de la Physique des
Plasmas) through their AMDA (Automated Multi-Dataset Analysis) and 3Dview tools
\citep{Genot:2017hp,Genot:2021cf}. 

The Cassini/RPWS/HFR instrument \replaced[id=R1]{records }{is recording} radio electric 
signals from 3.5 kHz to 16 MHz, using three electric antennas. This instrument allows 
\replaced[id=R1]{for reconstruction of }{to reconstruct} the absolute flux density and
polarization of the observed radio waves \citet{cecconi_RS_05}. In this study, we 
make use of two datasets from HFR: (a) the Cassini/RPWS/HFR Level 3e (N3e) dataset
\citep{https://doi.org/10.25935/9zab-fp47}, i.e., the ``Circular Polarization
Mode'' goniopolarimetric inversion as described in section 2.1.3.2 of
\citet{cecconi_RS_05}) ; and (b) the Cassini/RPWS/HFR SKR dataset 
\citep{https://doi.org/10.25935/zkxb-6c84} as described in \citet{lamy_JGR_08a}. 

The Cassini/MIMI instrument is designed for energetic neutral and charged particle
detection. MIMI consists of three sensors of which we use the Low Energy 
Magnetospheric Measurement System (LEMMS) sensor \citep{Krimigis:2004cf}. LEMMS is
a charged particle telescope with two units separated by 180$^{\circ}$ in pointing
that use solid state detectors and coincidence logic to determine the type of 
particle (electron or ion) and its energy, as well as magnetic deflection to
better separate ions from low energy ($\leq$800 keV) electrons
\citep{ROUSSOS2019}. 
In this study, we use data from the electron channels E4 with energy between 0.8 
and 4.7 MeV, with a \replaced[id=BC]{1-hr }{1 h} average window and non-background subtracted. 
\added[id=R2]{This 
channel shows MeV electron with a background level modulated by penetrating GCRs 
(Galactic Cosmic Rays). 
It was used to identify the Forbush decrease as a clear identification of the ICME 
\citep[see, ][]{Witasse:2017jv}.} \deleted[id=BC]{In addition, we use the proton channel P2 with 
energy 2.3 to 4.5 MeV from the calibrated Cassini/MIMI-LEMMS dataset} \added[id=BC]{The LEMMS 
data are }available at NASA/PDS \citep{PDS-PPI-Cassini-MIMI-LEMMS}. 

\begin{figure}
    \centering\includegraphics[width=\linewidth]{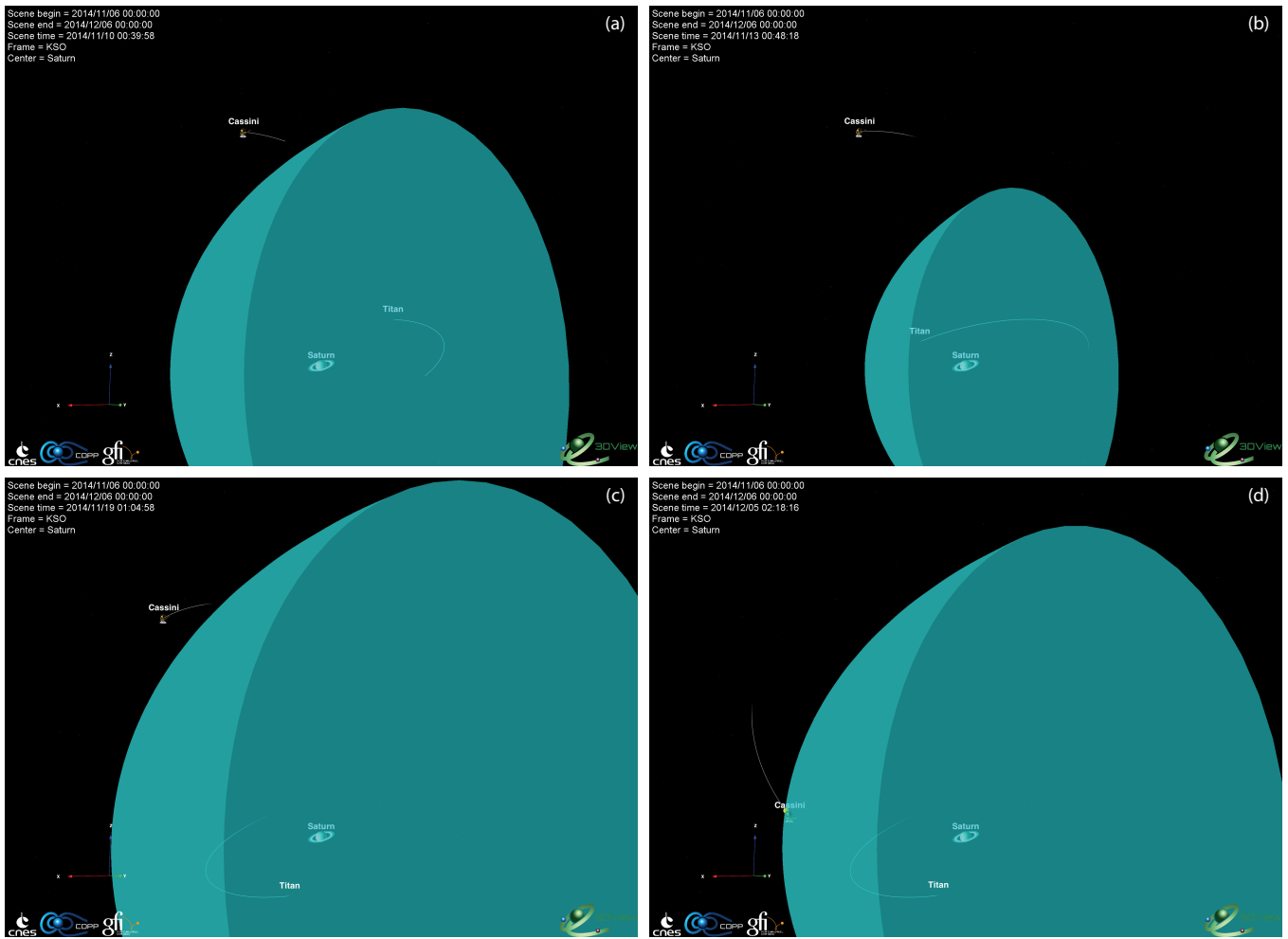}
    \caption{Cassini trajectory with Saturn's Magnetopause model 
    \cite{2010JGRA..115.6207K}. Snapshots from 3Dview, at 00:00 on days 2014-11-10,
    2014-11-13, 2014-11-19 and 2014-12-05, from (a) to (d) respectively. On panel 
    (d) a marker shows the modeled reentry into the magnetosphere.}
    \label{fig:3dview}
\end{figure}

\begin{figure}
    \centering
    \includegraphics{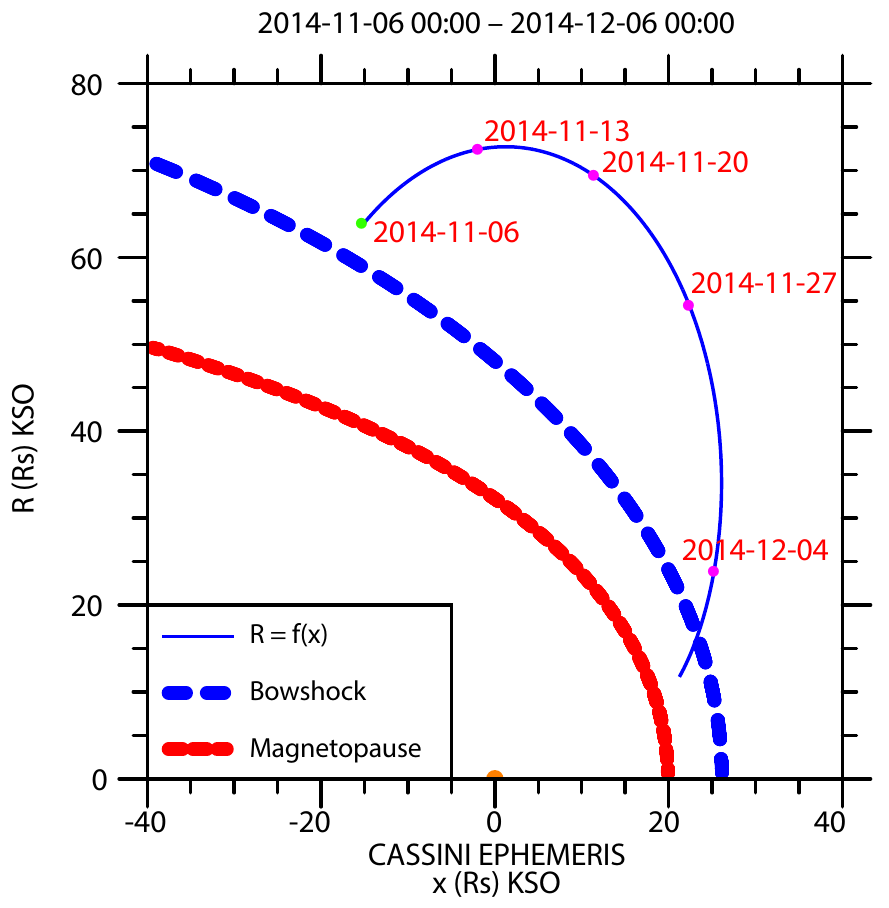}
    \caption{Overview of the Cassini spacecraft trajectory on KSO (Kronian 
    Solar Orbital) coordinates, as \replaced[id=CJ]{provided }{propided} by AMDA. The radial
    distance $R$ is displayed as a function of $x_\textrm{KSO}$, for the Cassini 
    trajectory and models of the \replaced[id=CJ]{magnetopause }{magnetosphere} and bowshock. 
    \added[id=CJ]{The modelled boundaries have been computed with a solar wind dynamic 
    pressure of 0.05 nPa. We expect this value to have increased significantly during the 
    ICME passage, which might explain why the boundary positions shown here don't necessarily 
    reflect the observed position of the boundaries. The pink and green dots along the trajectory
    are the location of Cassini at the labelled dates, the green location being the starting
    point.}}
    \label{fig:kso}
\end{figure}

The data have been studied during a 30-day interval (2014-11-06 00:00:00 to 
2014-12-06 00:00:00), \replaced[id=R1]{that includes the predicted }{including on the prediction} ICME 
arrival times at Saturn (see next sections). Figure \ref{fig:3dview} 
\replaced[id=R1]{shows }{is showing} an overview of the magnetospheric configuration, 
using 3DView \citep{Genot:2017hp}, and a dynamic magnetospheric model 
\citep{2010JGRA..115.6207K} implemented in this tool. The input \replaced[id=R1]{solar
wind}{Solar Wind} dynamic pressure is based on an MHD simulation 
propagation of OMNI data \citep{Tao:2005dp}, and is described in the next section 
(and Figure \ref{fig:enlil-amda}). Figure \ref{fig:kso} shows the Cassini trajectory 
together with two \replaced[id=BC]{models }{modeled}\added[id=BC]{: a} magnetopause \added[id=BC]{model} 
\citep{arridge_JGR_06} and \added[id=BC]{a} bowshock \added[id=BC]{model} \citep{Went:2011ge}, 
using a typical \replaced[id=R1]{solar wind }{ram} pressure of 0.05 nPa\added[id=CJ]{, which
corresponds to an average solar wind dynamic pressure, intermediate between compressed and 
rarefied conditions}. Those 
figures illustrate the geometrical configuration involved in this study: The 
Cassini spacecraft \replaced[id=R1]{is outside of }{comes out of} the Kronian magnetosphere 
at the beginning of the interval, on the Northern dawn side; it then travels in the
\replaced[id=R1]{solar wind }{Solar Wind}, while moving along the morning flank of the 
magnetosphere; it reenters the equatorial magnetosphere in the late morning local 
time at the end of the interval. 

\section{Observations and Models}
\label{sec:obs-models}
\begin{figure*}[p!]
    \centering\includegraphics[width=\textwidth]{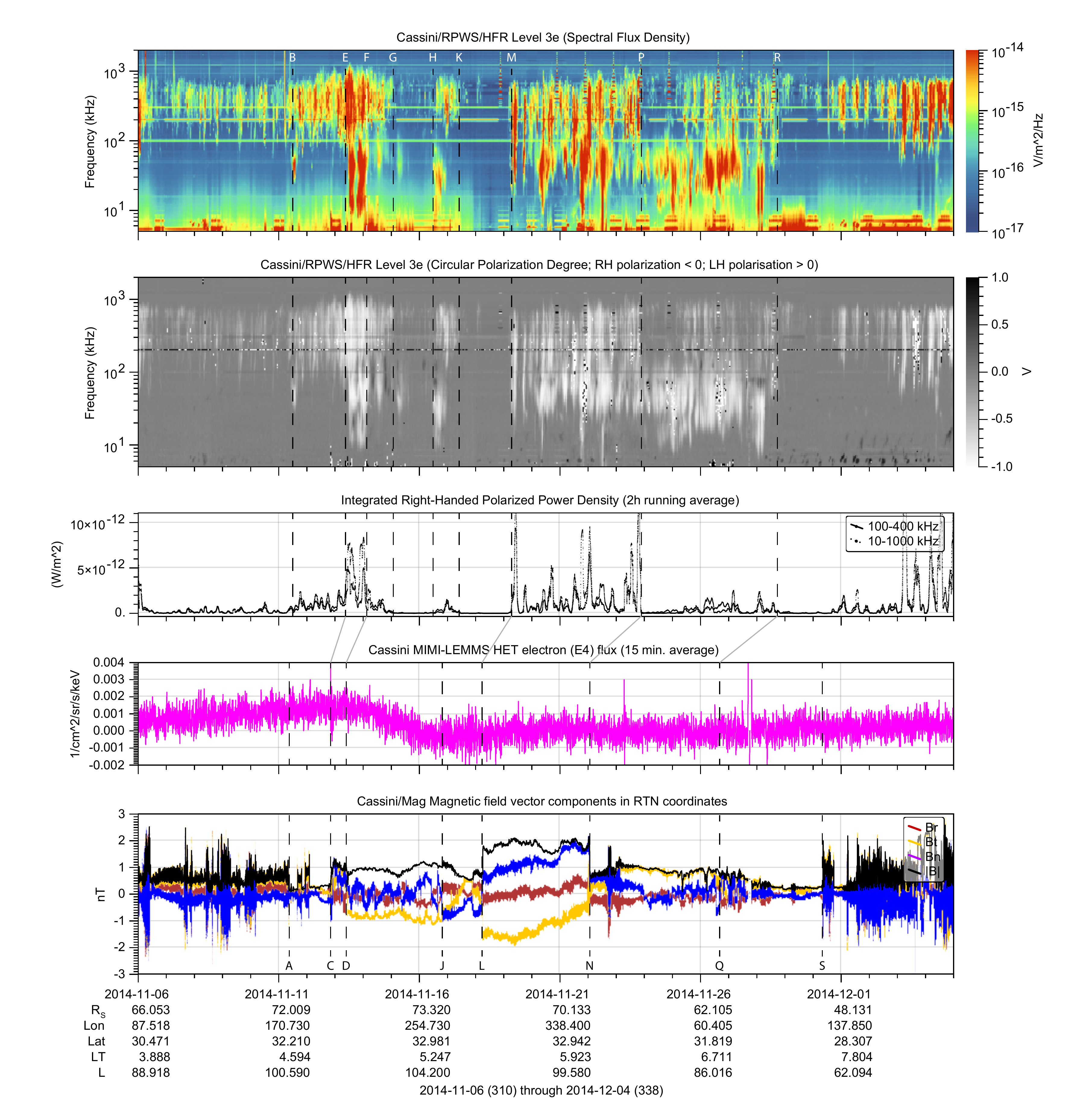}
    \caption{The from top to bottom: Radio electric flux density spectrogram; 
    Radio electric circular polarization degree spectrogram; SKR integrated flux 
    (plain line: 100-400 kHz range; dotted line: 10-1000 kHz range), with a 2 
    hours running average; energetic particle fluxes for electrons (magenta line) 
    \deleted[id=BC]{and protons (green line)} in units of $cm^{-2}sr^{-1}s^{-1}keV^{-1}$; and 
    magnetic field components in RTN coordinate system ($B_r$ in blue, $B_t$ in
    orange, $B_n$ in red, and $|B|$ in black). The tags (capital letters from A to 
    S) correspond to events described in Table \ref{tab:cme-skr timing}. Times 
    are given in the format 
    YYYY-MM-DD.}
    \label{fig:skr}
\end{figure*}

Figure \ref{fig:skr} \replaced[id=R1]{shows }{is showing} the radio electric and the 
\emph{in-situ} particle and magnetic field observations for the selected 30 days. 
The flux density and circular polarization degree spectrograms, and the SKR integrated 
flux time series are showing an intense SKR activity episode, including LFEs, 
from Nov.\ 11th to 29th, 2014:
\begin{itemize}
    \item \emph{Before Nov.\ 11th, 12:00}: The SKR is present with moderate 
    intensity, as expected for observations from early morning local time 
    \citep[$<$ 6hr LT, see, e.g., Figure 4 of][]{lamy_JGR_09}.
    \item \emph{From Nov.\ 11th, 12:00 to Nov.\ 15th 00:00}: A more intense 
    SKR episode is observed, with a pair of intense outburst accompanied with 
    LFEs on Nov.\ 13th at 12:00 and 19:30 (see Figure \ref{fig:skr-event-1}). 
    The LFE event is clear on the Integrated SKR Power panel, since the dashed 
    line (wide-band integration, 10 kHz to 1 MHz) is significantly higher than 
    the \replaced[id=CJ]{solid }{plain} line (medium-band integration, 100 to 400 kHz). This is also visible
    on the upper panel, where strong continuous emissions are observed between 10
    and 100 kHz.
    \item \emph{From Nov.\ 15th to Nov.\ 20th}: An SKR dropout is observed,
    with an isolated burst around Nov.\ 17th 00:00.
    \item \emph{From Nov.\ 20th to Nov.\ 24th}: Another intense SKR 
    episode is observed with strong and long lasting LFEs. 
    \item \emph{From Nov.\ 24th to Nov.\ 29th}: The LFE component is going on, 
    with moderate intensity SKR bursts.
    \item \emph{From Nov.\ 29th to Dec.\ 1st}: The SKR is dropping out and 
    resumes to expected levels at the end of the interval.
    \item \emph{From Nov.\ 29th to the end of the studied interval}: A 
    periodic very low frequency signal ($\sim$5 kHz) is observed, and mostly
    visible on the polarization panel. This feature corresponds to the 5 kHz
    Saturn narrow-band emission, so called n-SMR \citep{louarn_GRL_07} or 
    n-SKR \citep{wang_JGR_10}. 
\end{itemize}
The SKR dropouts observed in the morning sector of the magnetosphere is unusual, 
since this is the main activity LT sector of SKR 
\citep{galopeau_JGR_95,lamy_JGR_09}. 
Furthermore, one must keep in mind that the SKR modulation at or close to the 
planetary rotation period makes it difficult to accurately determine the actual 
onset of SKR events. The n-SMR feature observed at the end of the interval is 
typical of energetic magnetospheric events, as described in 
\replaced[id=R1]{detail }{details} in \citet{louarn_GRL_07}. Their Figure 1 shows similar
LFEs followed by n-SMR pulsed emissions. The \emph{in-situ} data (magnetic field 
and particle) displays several characteristic signatures which will be described 
in section \ref{sec:icme}. 

\begin{figure}[h]
    \centering\includegraphics[width=0.7\linewidth]{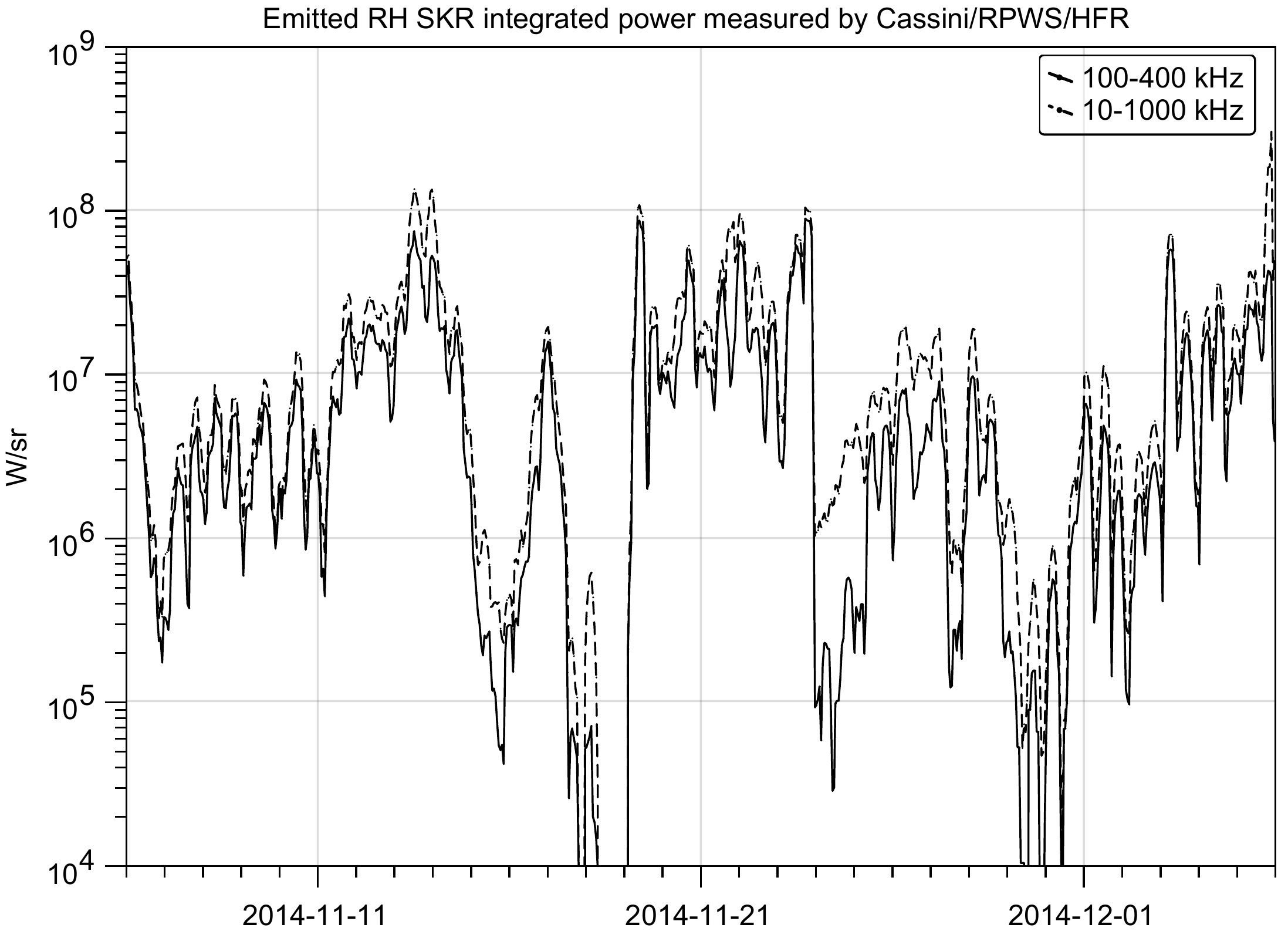}
    \caption{Right Handed polarized emitted SKR power integrated on 100-400 kHz 
    (plain line) and 10-1000kHz (dashed line), as measured by Cassini/RPWS/HFR. 
    The data has been averaged over 15 minutes.}
    \label{fig:skr-log}
\end{figure}

The variability of the emitted SKR power \citep{lamy_JGR_08a} 
is presented on Figure \ref{fig:skr-log}. The average usual SKR emitted power 
is of the order of $10^6$ W/sr to $10^7$ W/sr, and we observe SKR average 
power raising up to $10^7$ W/sr, with peaks to $10^8$ W/sr (i.e., one order of 
magnitude higher than regular levels). SKR emitted power \replaced[id=R1]{drops }{is dropping} 
down to $10^5$ W/sr and even down to undetectable levels on day Nov.\ 18th, 2014, 
and on the last two days of Nov.\ 2014.

\begin{figure}[h]
    \centering\includegraphics[width=\linewidth]{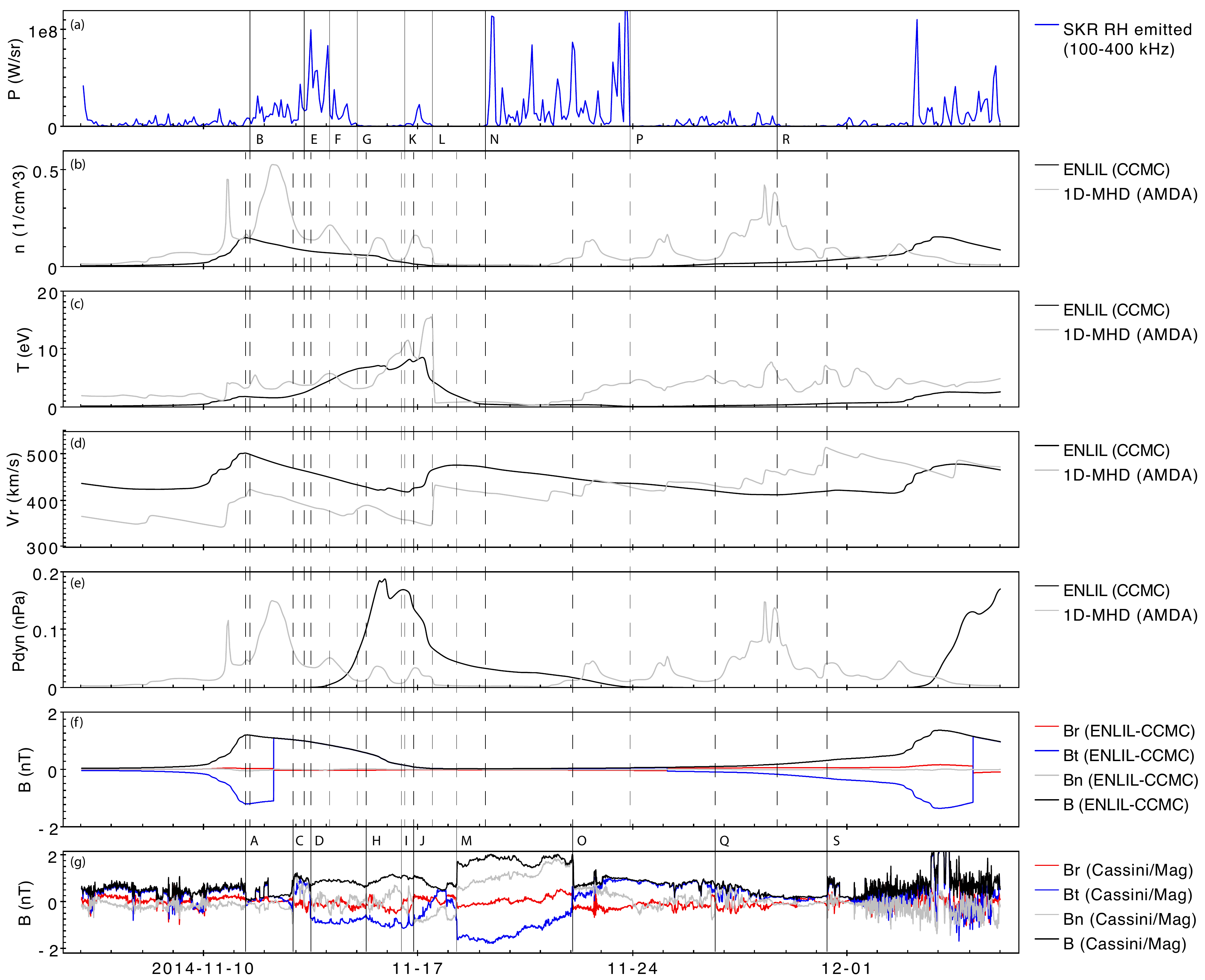}
    \caption{Comparison of the solar wind modeled parameters from the two 
    selected propagation models. From top to bottom: (a) the SKR RH emitted
    power density integrated over 100-400 kHz and smoothed to a temporal 
    resolution of 6000 seconds; (b) to (e) the modelled solar wind density (in 
    cm$^{-3}$), temperature (in eV), radial velocity (in km/s) and dynamic
    pressure (in nPa), as provided by the ENLIL code at CCMC (black) and 
    1D-MHD code at CDPP (grey); (f) and (g) the modelled (ENLIL) and measured 
    (Cassini/Mag) solar wind magnetic field in RTN cooordinates ($B_r$ in 
    red; $B_t$ in blue; $B_n$ in grey; $B$ magnitude in black). The vertical 
    lines and the capital letter markers are the same as in Figure 
    \ref{fig:skr} and described in Table \ref{tab:cme-skr timing}. Times are 
    given in the format YYYY-MM-DD.}
    \label{fig:enlil-amda}
\end{figure}

The propagated \replaced[id=R1]{solar wind }{Solar Wind} data available in AMDA are presented in Figure 
\ref{fig:enlil-amda}. This Figure shows the modeled \replaced[id=R1]{solar wind }{Solar Wind} \replaced[id=BC]{density, temperature, radial velocity and dynamic pressure (panels (b) to (e)) }{dynamic pressure (top 
panel)}, using the 1D MHD model developed by \citet{Tao:2005dp}. Setting a 
threshold to 0.1 nPa, AMDA provides us with four time intervals of enhanced 
dynamic pressure, with peak values at about 0.15 nPa. The four intervals are 
shown in Table \ref{tab:amda-tao-peak}. The model is however not well 
constrained \replaced[id=BC]{since the simulation is run with input parameters 
measured at Earth, which in Solar conjunction with Saturn at this time. }{as we can see on the lower panel of Figure \ref{fig:tao}, 
since Saturn is in Solar conjunction from Earth.} \added[id=BC]{This model shall thus 
only be used as an rough indication of the solar wind conditions.}

\begin{table}[h]
    \centering
    \begin{tabular}{l|l|l}
         ID & Start Time & End Time \\
         \hline
         1  & 2014-11-10 17:55 & 2014-11-10 19:25 \\
         2  & 2014-11-11 20:55 & 2014-11-12 17:25 \\
         3  & 2014-11-28 06:05 & 2014-11-28 09:05 \\
         4  & 2014-11-28 12:35 & 2014-11-28 18:35
    \end{tabular}
    \caption{AMDA derived dynamic pressure peaks ($>$ 0.1 nPa) as modeled by 
    \citet{Tao:2005dp}. Times are given in the format YYYY-MM-DD hh:mm}
    \label{tab:amda-tao-peak}
\end{table}


Propagated \replaced[id=R1]{solar}{Solar} wind data available at CCMC (Community Coordinated Modeling 
Center) are also presented in Figure \ref{fig:enlil-amda}. The simulation uses ENLIL, a 
3D MHD modeling code \citep{ODSTRCIL2003497}. The simulation run used here is 
\texttt{Leila\_Mays\_092716\_SH\_1}\footnote{Available at:
\url{https://ccmc.gsfc.nasa.gov/database_SH/Leila_Mays_092716_SH_1.php}} 
\citep{Witasse:2017jv}. \added[id=BC]{The Figure presents the same solar wind
parameters, as for the 1D-MHD model presented in the previous paragraph, as well 
as the modelled magnetic field components.} The Figure shows a clear enhancement 
of dynamic pressure from Nov.\ 15th to Nov.\ 17th, with a peak value around 0.2 nPa. 
The ENLIL data also shows another similar event starting at the end of the 
time interval, on Dec 4th. 

The two propagation models are predicting \added[id=BC]{similar 
solar wind parameters values and trends (density, temperature and radial velocity) 
up to $\sim$Nov.\ 21st, with the noticeable exception of the dynamic pressure,
which is showing different peak times depending on the model, as well as rather
different overall variation all along the studied interval. The modelled magnetic 
field doesn't reproduce the observed ICME structure.} \added[id=CJ]{During the 
interval when Cassini exited into the solar wind (19 days, from labels A to S of 
figure \ref{fig:skr} and Table \ref{tab:cme-skr timing}), it was possible to 
directly sample the interplanetary magnetic field. We can use this to compare 
against model predictions for the broader interval.} \added[id=BC]{The two models 
are predicting} different arrival times of the ICME. \replaced[id=R1]{The model 
developed by \mbox{\citet{Tao:2005dp}} and available from the CDPP }{The 
\mbox{\citep{Tao:2005dp}} model from CDPP} is about 5 days \replaced[id=R1]{earlier 
}{early} compared to the ENLIL simulation run output from CCMC. Due to the solar 
system orbital configuration, the ENLIL simulation run should be more accurate. 
\added[id=OW]{Indeed, the ENLIL simulation predicts the arrival of the ICME at 
12:00 UT on 15 November, about 2.5 days after the real hit. This is a rather 
accurate estimation given the distance of Saturn with respect to the Sun and 
the limitations of the simulation at these distances \citep{Witasse:2017jv}}. 

\section{ICME Event}  
\label{sec:icme}
In order to correlate and interpret the variations of the SKR emissions, it 
is necessary to \replaced[id=R1]{characterise }{draw a picture of} the ICME event. This ICME erupted from the 
Solar Active Region 12192 on 14 October 2014 at $\sim$18:30 UT as seen by the 
SOlar and Heliospheric Observatory (SOHO), the Project for On Board Autonomy 2 
(PROBA-2), the Solar Dynamics Observatory (SDO), and the Solar 
TErrestrial RElations Observatory Ahead (STEREO-A) missions. \replaced[id=R1]{On its transit 
to the outer solar system, the ICME encountered Venus, STEREO-A, Mars, comet 
67P/Churyumov-Gerasimenko, Saturn, }{In its way towards the outer 
solar system hit STEREO-A, Venus, Mars, comet 67P/Churyumov-Gerasimenko, 
Saturn, and} possibly New Horizons on its way to Pluto and Voyager-2 at the 
heliosheath as discussed in detail by \citet{Witasse:2017jv}. 
The propagation of this event was also modelled with two different solar wind
simulations\replaced[id=R1]{: }{, such as} \added[id=BC]{(i)} the CDPP Propagation Tool, and \added[id=BC]{(ii)}  the Wang-Sheeley-Arge 
(WSA)-ENLIL + Cone model. For this \replaced[id=BC]{latter }{latest} case and based on the Graduated 
Cylindrical Shell (GCS) fit performed to the coronagraph observations from 
STEREO-A and SOHO, the CME was injected into the simulation using \deleted[id=R1]{as} inputs 
\added[id=R1]{of} 1015 km/s \replaced[id=R1]{for }{of} radial speed, 150$^{\circ}$ \replaced[id=R1]{for }{of} 
longitude, 12$^{\circ}$ \replaced[id=R1]{for }{of} latitude
and a full width of 116$^{\circ}$. For the ICME propagation out to Saturn, a 
medium resolution (2$^{\circ}$) simulation was performed on a HEEQ spherical 
grid of 1920${\times}$60${\times}$180 (r, $\theta$, $\phi$) \deleted[id=R1]{size}, with a range 
of 0.1 to 10.1 AU in radius (r), -60$^{\circ}$ to +60$^{\circ}$ in latitude 
($\theta$), and -180$^{\circ}$ to 180$^{\circ}$ in longitude ($\phi$). In 
addition to the CME of this study, the simulation also included 138 CMEs with 
speeds above 500 km/s and full angular widths above 50$^{\circ}$. \added[id=R1]{This was 
necessary in order to get the most accurate estimation of the arrival time of the ICME 
to Saturn since most CMEs merge with the background solar wind at largest distances}
\added[id=CJ]{\citep[see, e.g., ][]{2004JGRA..109.9S03H}}. Figure 
\ref{fig:enlil_still} shows a \replaced[id=R1]{snapshot }{still} of the propagation of this ICME 
just before its arrival at Saturn (the ICME is outlined in black), and Figure
\ref{fig:enlil-amda} shows the derived solar wind density, velocity and temperature 
profiles obtained from the simulation at Saturn. The ICME arrived at Saturn
approximately one month after its ejection at the Sun as observed by Cassini, 
which was immersed in the solar wind at this time. 

Although there were no \emph{in-situ} \added[id=R1]{plasma} measurements of the ICME 
and the background 
solar wind speeds, values of 500 and 400 km/s respectively can be estimated 
from \citet{Witasse:2017jv}. This estimation is consistent with the modelled 
\replaced[id=R1]{solar wind }{Solar Wind} velocity presented on Figure \ref{fig:enlil-amda}.

Table \ref{tab:cme-skr timing}
gives a summary of the timing of the different regions of the ICME at Cassini 
as well as the corresponding SKR enhancements. \added[id=R2]{In particular, the 
shock of the CME is seen as a significant increase of the interplanetary magnetic 
field strength from 0.3 nT to 1.2 nT. It is followed by the sheath/ejecta, where a 
clear rotation of the $B_y $component is seen. After that, there is the magnetic 
cloud that is clearly identified with a rotation of the magnetic field $B_z$ 
component, with the addition that this event has a ``magnetic bottle'' structure 
which is identified as the region where the magnetic field strength is at its 
largest value (1.6-2 nT).} \added[id=CJ]{The internal structure of a magnetic bottle is comprised
of helical, twisted magnetic field lines, which present as a local intensification
of the field strength when traversed by a spacecraft.} 

\begin{table}[h]
    \centering
    \hspace*{-1.5cm}
    \begin{tabular}{l|c|l|l|ll}
         Event                      & Tag &  ICME              & SKR               & \multicolumn{2}{l}{Delay (hr)}  \\
         \hline
         Magnetosheath (outwards)     & A &  \multicolumn{2}{c|}{2014-11-11 09:00} & --       \\
                                      & B & --                 & 2014-11-11 12:00  & --       \\
         Shock arrival                & C & 2014-11-12 22:50   & --                & --       \\
         Sheath/ejecta                & D & 2014-11-13 11:30   & --                & --       \\
                                      & E & --                 & 2014-11-13 11:40  & 12.8 & from C \\
                                      & F & --                 & 2014-11-14 03:00  & 15.5 & from D \\
                                      & G & --                 & 2014-11-15 02:00  & 38.5 & from D \\
         \deleted[id=BC]{Proton injection (start)}     & \deleted[id=BC]{H} & \deleted[id=BC]{2014-11-15 09:30}   & \deleted[id=BC]{--} & \deleted[id=BC]{--} \\
         \deleted[id=BC]{Proton injection (end)}       & \deleted[id=BC]{I} & \deleted[id=BC]{2014-11-16 10:30}   & \deleted[id=BC]{--} & \deleted[id=BC]{--} \\
                                      & \replaced[id=BC]{H }{J} & --                 & 2014-11-16 12:00  & \replaced[id=BC]{-- }{26.5} & \replaced[id=BC]{-- }{from H} \\
         Magnetic cloud start         & \replaced[id=BC]{J }{K} & 2014-11-16 19:30   & --                & --       \\ 
                                      & \replaced[id=BC]{K }{L} & --                 & 2014-11-17 11:00  & \replaced[id=BC]{-- }{24.5} & \replaced[id=BC]{-- }{from I} \\
         Magnetic bottle (start)      & \replaced[id=BC]{L }{M} & 2014-11-18 06:10   & --                & --       \\ 
                                      & \replaced[id=BC]{M }{N} & --                 & 2014-11-19 07:20  & 25.2 & from \replaced[id=BC]{L }{M} \\ 
         Magnetic bottle (end)        & \replaced[id=BC]{N }{O} & 2014-11-22 01:30   & --                & --       \\ 
                                      & P & --                 & 2014-11-23 20:10  & 42.7 & from \replaced[id=BC]{N }{O} \\
         Magnetic cloud end/sheath    & Q & 2014-11-26 19:30   & --                & --       \\ 
                                      & R & --                 & 2014-11-28 19:00  & 47.5 & from Q \\ 
         Magnetosheath (inwards)      & S & \multicolumn{2}{c|}{2014-11-30 08:40}  & --       \\

         \end{tabular}
    \caption{Timings of the major signatures of the ICME and SKR. The first 
    column are the tags shown on Figure \ref{fig:skr}. Times are given in the 
    format YYYY-MM-DD hh:mm}
    \label{tab:cme-skr timing}
\end{table}


\begin{figure}[p]
    \centering
    \includegraphics[width=\linewidth]{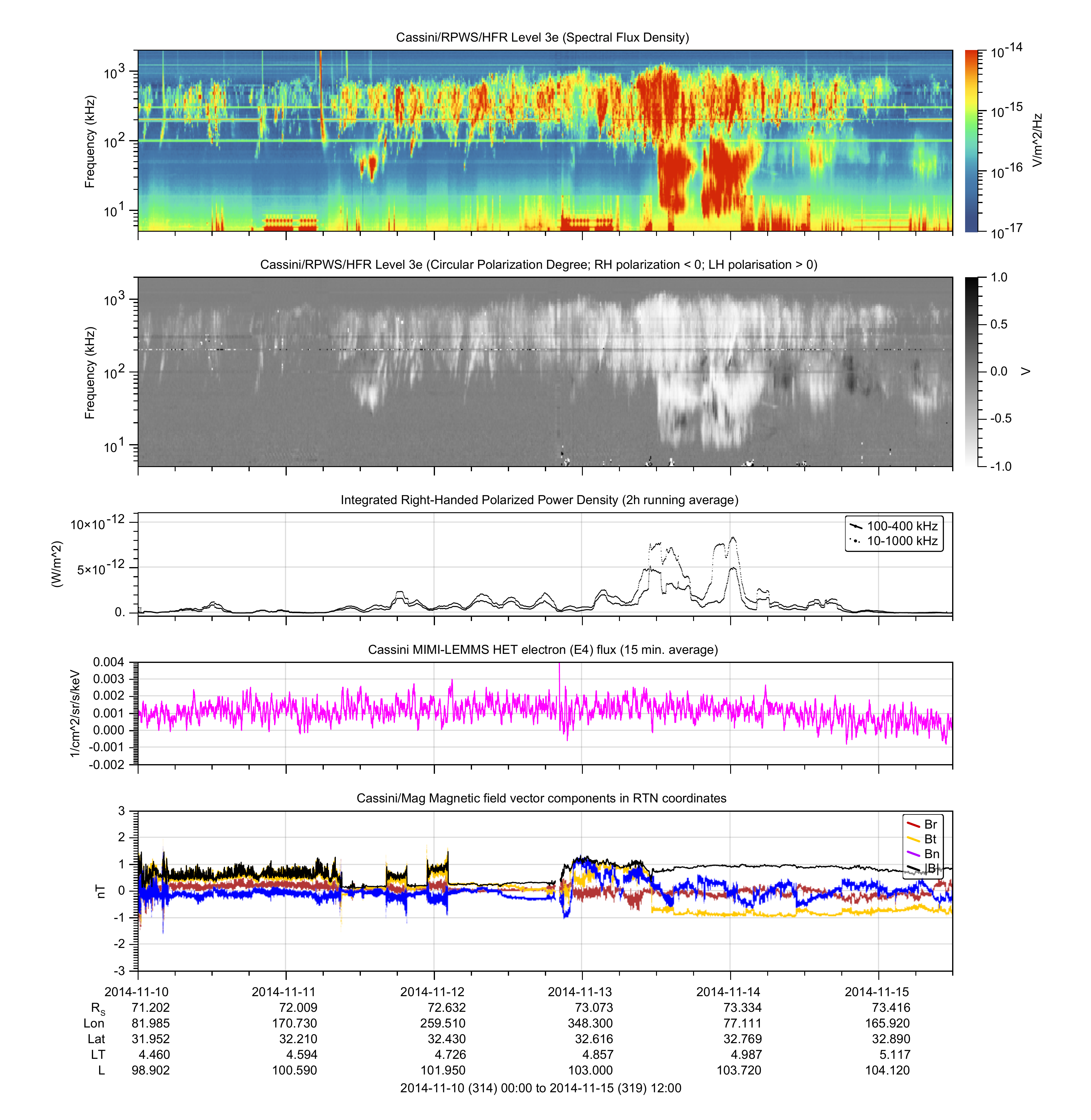}
    \caption{Zoomed-in version of Figure \ref{fig:skr}, from Nov.\ 10th 
    00:00 SCET to Nov.\ 15th, 12:00 SCET. The panel description is the 
    same as for Figure \ref{fig:skr}. Times are given in the format 
    YYYY-MM-DD, and hh:mm.}
    \label{fig:skr-event-1}
\end{figure}

\begin{figure}[p]
    \centering
    \includegraphics[width=\linewidth]{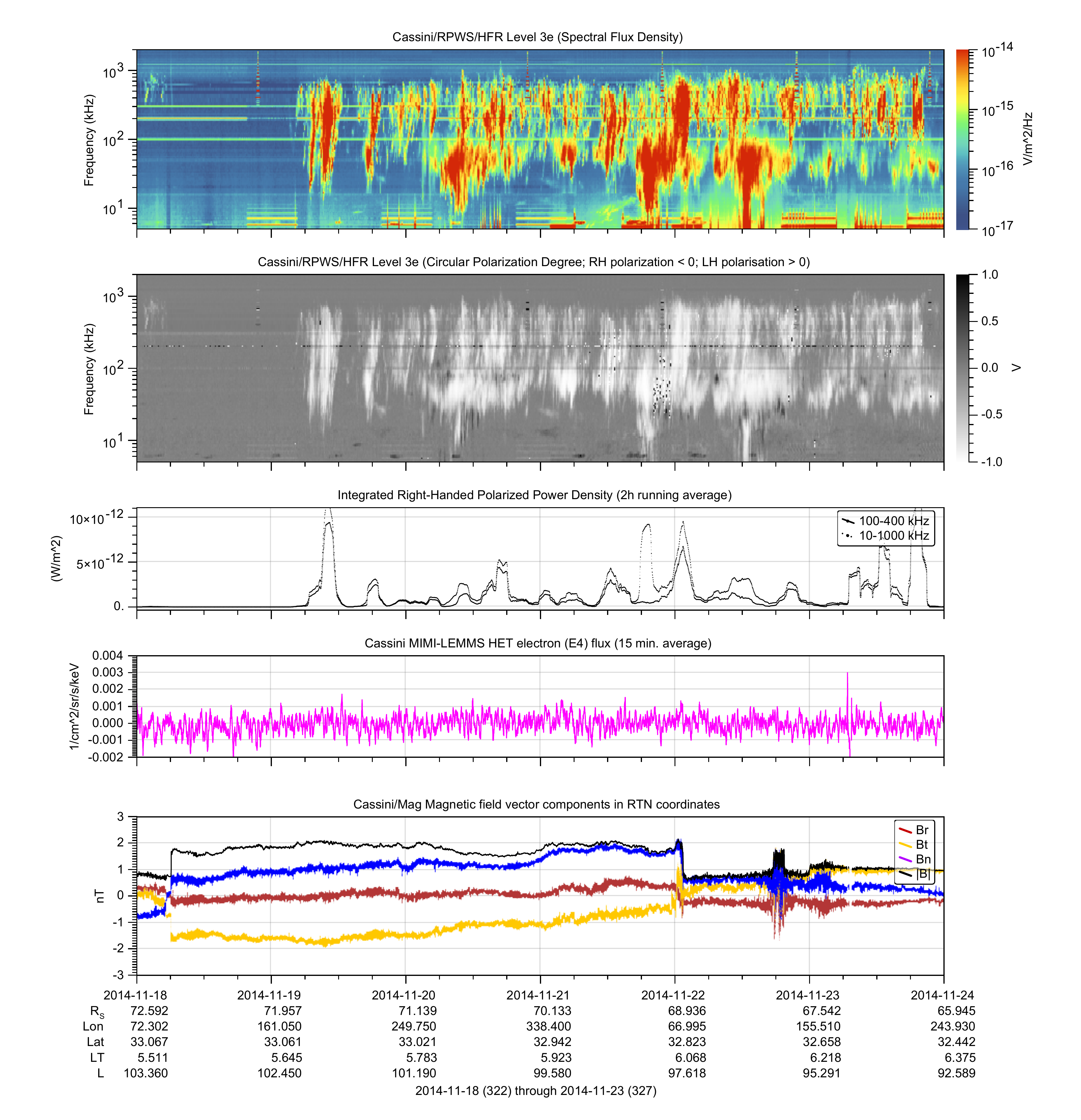}
    \caption{Zoomed-in version of Figure \ref{fig:skr}, from Nov.\ 18th 
    00:00 SCET to Nov.\ 24th 10:00 SCET. The panel description is the 
    same as for Figure \ref{fig:skr}. Times are given in the format 
    YYYY-MM-DD, and hh:mm.}
    \label{fig:skr-event-2}
\end{figure}


It is noticeable that the enhanced activity of the SKR starts with the \replaced[id=R1]{arrival
of the first event in the ENLIL simulation }{first ENLIL simulation run event}, during the 
enhanced \replaced[id=R1]{solar wind }{Solar Wind} density event, before 
the peak dynamic pressure. The isolated SKR burst emission on Nov.\ 17th (during 
phase (c)) corresponds to the peak of dynamic pressure predicted by the ENLIL 
simulation run. The rise of SKR activity \added[id=R1]{at} the end of the interval also 
\replaced[id=R1]{corresponds }{correspond} \replaced[id=R1]{to a period of enhanced density and dynamic 
pressure in the ENLIL simulation }{to the ENLIL enhanced density (and dynamic pressure)}.

\section{Discussion}

The SKR variability observed the month of Nov.\ 2014 is closely related to the 
ICME events, with observed delays of $\sim$ 20h between SKR and \replaced[id=R1]{solar 
wind }{Solar Wind} events. 
We observe that the integrated power (over 100-400 kHz, i.e, the core of the 
SKR band, as well as over the 10-1000 kHz, i.e., the extended SKR band) increase 
by 1 order of magnitude. We also observe SKR dropouts at undetectable levels. 
\added[id=BC]{With the assumption of a direct control of the SKR intensity by 
the solar wind dynamic pressure \citep{kaiser_Sci_80},} 
\replaced[id=BC]{such }{Such} low SKR level conditions are indicating significant 
expansion of the Kronian magnetosphere (or, equivalently, a strong depletion of 
the \replaced[id=R1]{solar wind}{Solar Wind}). It is noticeable that 
\replaced[id=R1]{these trends are }{the increase, dropout and increase again 
is} very similar as the event presented in Figure 1 of \citet{jackman_JGR_05}. 
\added[id=BC]{\citet{desch_JGR_83} reported the disappearance of SKR when 
Saturn is in Jupiter's magnetotail, hence in a depleted plasma environment.}

The spacecraft is located on the dawn side of the magnetosphere ($\sim$0700 hr 
LT) \added[id=CJ]{throughout this interval}. \replaced[id=CJ]{It is expected due
to the location of radio sources and the hollow cone beaming pattern of the SKR
that radio emissions in the main SKR band should maximise around this viewing 
location \mbox{\citep{lamy_JGR_08a}}. Indeed, Figure \ref{fig:skr} displays 
continuous SKR modulations around Nov.\ 11th 2014, as expected from this vantage 
point.}{This local time sector should display 
continuous SKR modulations, as 
observed before Nov.\ 11th, 2014 on Figure \ref{fig:skr}, as described in
\mbox{\citet{lamy_JGR_08a}}} The SKR modulation breakouts, as well as SKR peaks 
and dropouts are \replaced[id=BC]{thus }{clearly} \added[id=R1]{related to} 
a specific set of events, responding to the dynamics, rather than an effect 
of limited viewing due to spatial constraints.

The intensification in the main band of SKR is accompanied by an extension of 
the emission to lower frequencies \added[id=CJ]{(LFE)}, similarly to what is observed in the case
of a CIR \added[id=CJ]{compressions} impacting the Kronian magnetosphere, as investigated in \replaced[id=R1]{detail }{details} by
\citet{jackman_JGR_05}. This strong, continuous extension of emission is 
interpreted as a growth/expansion of the radio source to higher 
altitudes along the field line, with the lower field strength correlated to 
lower frequency emission (due to the \replaced[id=BC]{CMI }{cyclotron maser instability} generation 
mechanism). 

We observe an enhancement \replaced[id=BC]{of SKR intensity after \emph{in-situ} events 
}{in the SKR $\sim$20 hours after \emph{in-situ} events,} 
with delays ranging from $\sim$13 hr \replaced[id=R1]{after }{at} the shock arrival, to $\sim$25 hr
at the \deleted[id=BC]{proton injection and} \added[id=BC]{start of the} magnetic bottle \deleted[id=BC]{start}, and to $\sim$44 hr 
near the end of the event. The uncertainty on the delays shall be evaluated
taking into account the SKR modulation at or close to the 
planetary rotation period, i.e., $\sim$10.5 hr \citep{Lamy:2010vz}. \replaced[id=BC]{Considering a 
typical SKR burst duration of $\sim$5.2 hr (half the planetary rotation period), the uncertainty is asymmetric: 
from about $-$5.2 hr (if the event occurs during the low SKR level modulation phase) to 0 hr (if the 
event occurs during the high SKR level modulation phase). }{This 
implies an asymmetric uncertainty from 0 (if the event occurs during the 
high SKR level modulation phase) to $\sim$5.2 hr (1/2 of the modulation
period, if the event occurs during the low SKR level modulation phase).} 
\added[id=BC]{This implies that delays can be overestimated by a few hours.}
Our findings on delays between \emph{in-situ} and radio events are fully consistent 
with the study of \citet{Taubenschuss06}, where delays between \replaced[id=R1]{solar 
wind }{Solar Wind} and SKR events have been derived using statistical methods. This previous 
study identified delays ranging from $\sim$13 hr for dynamic pressure 
enhancement, to $\sim$27 hr to $\sim$44 hr for magnetic field events.  
\added[id=BC]{The observed delays can be expressed in units of the average planetary 
rotation period: 13 hr, 25 hr and 44 hr correspond respectively to 1.2, 2.4 and 4.2 
planetary rotation periods with an uncertainty of $-$0.5/$+$0 planetary rotation. \citet{jackman_JGR_05}
reported delays of about 40 hours. \citet{Rucker:1984ws} reported delays of  
a few planetary rotation, with a maximum correlation at about 4 planetary rotations. 
Interestingly, Figure 6 of \citet{Rucker:1984ws} shows a superposed-epoch analysis of 
SKR peaks with various solar wind parameters: a statistical delay of a few planetary 
rotations is observed (the figure is too small to get a better estimate). This suggests 
that the delay in the response of SKR to solar wind variability is always present. In the 
general case, the SKR periodic variability makes it difficult to evaluate the delays.
However, since the event studied here triggers a drastic change in SKR properties, the 
delay can be estimated directly.}

\added[id=BC]{Such long delays are still not understood. \citet{jackman_auroral_2013} 
reported substorm-like event in Saturn's magnetotail at a distance of $\sim$45 
$R_s$. With the modelled solar wind velocity (400 to 500 km/s, see Figure 
\ref{fig:enlil-amda}), the solar wind event is reaching this distance after a 
few hours at most. The observed delays are thus inconsistent with a direct control
by the solar wind. The variability of SKR is modulated by several factors: an active 
local sector \citep{galopeau_JGR_95}, which location is driven by the the solar wind 
velocity \citep{galopeau_variations_2000} and the rotating magnetospheric plasma (see,
e.g., \citet{Lamy:2013em}, which describes in detail the relation between the 
magnetospheric plasma rotation (observed with an energetic neutral atoms imager) and 
the auroral activity (observed at radio, infrared and ultraviolet spectral ranges). 
As shown in that latter study, the dynamics and the transport of plasma in the 
magnetosphere are key drivers of the auroral variability.}

\added[id=BC]{According to the CMI mechanism, several parameters can 
play a role in the enhancement or damping of radio emissions: the growth rate is 
related to the presence of free energy (unstable particle distribution functions, 
such as energetic beams, loss cone or shell distributions), as well as a locally 
depleted and magnetized cold plasma. SKR bursts and 
dropouts are observed during the event studied here. This indicates a radical 
reconfiguration of the magnetosphere, resulting in a change of the plasma 
properties in the radio source regions (i.e., along the auroral magnetic field lines). 
Understanding the magnetospheric response to intense solar wind events requires 
global and dynamical modelling of Saturn's magnetosphere at scales of a few planetary
rotations.}

Thanks to the specific orbital configuration during the event, we have 
been able to observe simultaneously the ICME in the \replaced[id=R1]{solar wind }{Solar 
Wind} upstream
from Saturn, as well as the SKR response. This study is thus very 
complementary to other studies \citet[see, e.g.,][]{Palmerio:2021ib} 
using only modelled data to identify the event arrival times at Saturn.  
The shock arrival time derived from the data is estimated to 
2014-11-12 22:50. The 1D-MHD modelled data \citep{Tao:2005dp} predicts a 
first dynamic pressure pulse starting at 2014-11-10 17:55, and a second, 
longer one, at 2014-11-11 20:55. The ENLIL model \citep{ODSTRCIL2003497}
predicts a peaked solar wind density on 2014-11-11 and a peaked 
dynamic pressure on 2014-11-15. In terms of time of arrival estimations, 
our observational data seem to favor the 1D-MHD modelling available in 
AMDA provides, compared to the ENLIL code available at CCMC. 
\added[id=BC]{However, it is clear that current solar wind modelling tools are 
not providing solar wind parameters with sufficient timing accuracy for 
detailed comparison with \emph{in-situ} measurement at the distance of Saturn.}

\section*{Conflict of Interest Statement}
The authors declare that the research was conducted in the absence of any commercial 
or financial relationships that could be construed as a potential conflict of interest.

\section*{Author Contributions}
BC performed writing, conceptualization, data gathering, figure editing. 
OW, CMJ and BSC performed conceptualization, writing -- reviewing and editing. 
MLM provided CCMC data.


\section*{Funding}
BC acknowledges support from Observatoire de Paris, CNRS/INSU (Centre National de 
la Recherche Scientifique / Institut des Sciences de l'Univers) and CNES (Centre 
National d'Etudes Spatiales). OW was supported by ESA (European Space Agency). 
CMJ's work at DIAS was supported by the Science Foundation Ireland Grant 18/FRL/6199. 
BSC acknowledges support through UK-STFC Ernest Rutherford Fellowship 
ST/V004115/1 and STFC grants ST/S000429/1, \added{ST/W00089X/1}.

\section*{Acknowledgments}
The authors thank the developers of: AMDA (Automated Multidataset Analysis Tool) 
and 3Dview at CDPP (Centre de Donn\'ees de la Physique des Plasmas); Autoplot at 
University of Iowa; and TOPCAT (Tools for Operations on Catalogues and Tables) at 
Bristol. Figures \ref{fig:skr}, \ref{fig:skr-log}, \ref{fig:skr-event-1} and 
\ref{fig:skr-event-2} have been produced with 
Autoplot \citep{Faden:2010jo}. The Cassini/RPWS data displayed on the figures have 
been retrieved from PADC (Paris Astronomical Data Centre) using the \textit{das2} 
\citep{piker_2017} interfaces of the MASER (Measurements, Analysis, and Simulation 
of Emission in the Radio range) team \citep{Cecconi:2020bc}. The Cassini/MAG is 
distributed by University of Iowa, also through a \textit{das2} interface. Figure 
\ref{fig:enlil-amda} have produced with TOPCAT \citep{Taylor:2005wq}, after 
conversion of the ENLIL original tabular data into VOTable \citep{votable}, and
transferring the other datasets from AMDA \citep{genot_ASR_10,Genot:2021cf} using 
SAMP (Simple Application Messaging Protocol) \citep{2014A&C.....7...62G,2015A&C....11...81T}.
\added{The authors thank the two reviewers for their helpful comments, as well as
L.\ Lamy and V.\ Génot, who sent valuable comments based on the initial 
manuscript version published on ArXiv.org.}

\section*{Data Availability Statement}
The datasets analyzed for this study and their location have been described 
the section \ref{sec:data-traj}. \added[id=BC]{Appendix \ref{sec:supp-mat} 
presents material, which can be used to reproduce figures of this paper.}

\appendix
\section*{Appendix}
\section{Supplementary material}
\label{sec:supp-mat}
The supplementary material for this paper \citep{supplementary} is available from \url{https://doi.org/10.25935/DZRB-P221}, and contains the following content.

\subsection{Autoplot configuration files}
Three Autoplot configuration files (\texttt{.vap} extension) are available:
\begin{itemize}
    \item \texttt{icme-skr-fig-full.vap} to reproduce Figure \ref{fig:skr}.
    \item \texttt{icme-skr-fig-skr-emitted.vap} to reproduce Figure \ref{fig:skr-log}.
    \item \texttt{icme-skr-fig-event-1.vap} to reproduce Figure \ref{fig:skr-event-1}.
    \item \texttt{icme-skr-fig-event-2.vap} to reproduce Figure \ref{fig:skr-event-2}.
\end{itemize}
\subsection{3Dview files}
\begin{itemize}
    \item \texttt{icme-skr-3dview.mov}: 3Dview \citep{Genot:2017hp} generated
    movie, showing the Cassini trajectory during the studied interval. The 
    Cassini/MAG magnetic field vector is plotted along the trajectory, with a 
    rainbow color map for the magnetic field amplitude. The received SKR RH 
    integrated power time series (on the 10 to 1000 kHz band) is also plotted 
    along the trajectory, in the orbit plane, with a white-blue color map for 
    the power values. The magnetopause location is also displayed, using the
    \citep{2010JGRA..115.6207K} model, with the dynamic pressure input provided 
    by 3dview from \citep{Tao:2005dp}.   
    \item \texttt{icme-skr-3dview.3dv}: 3Dview configuration file to reproduce 
    \texttt{icme-skr-3dview.mov}. 
\end{itemize}
\subsection{TOPCAT files}
\begin{itemize}
    \item \texttt{ENLIL-AMDA.xml}: A TOPCAT session VOTable file containing 
    several tables: (i) the ENLIL solar wind parameters from the CCMC modelling 
    run; (ii) the 1D-MHD propagated solar wind parameters \citep{Tao:2005dp}; 
    (iii) the SKR emitted power of its RH component, from the Cassini/RPWS/SKR 
    dataset \citep{https://doi.org/10.25935/zkxb-6c84}, with a temporal 
    resolution of 600 seconds; (iv) the same dataset with a resolution of 
    6000 seconds; and (v) the measured magnetic field \citep{Dougherty:2004tr}, 
    with a 60 seconds temporal resolution. Datasets (ii) to (v) have been 
    transferred from AMDA to TOPCAT thanks to the SAMP protocol
    \citep{2014A&C.....7...62G,2015A&C....11...81T}. Upon loading this file 
    into TOPCAT, a set of predefined data selections and computed variables will 
    be available to the user. 
    \item \texttt{ENLIL-AMDA.fits}: The same content as in the previous item, 
    exported from TOPCAT using the FITS format option.
    \item \texttt{ENLIL-AMDA.txt}: A STILTS script that can be used to reproduce
    Figure \ref{fig:enlil-amda}.
    \item \texttt{README.txt}: Details on how to use these files. 
\end{itemize}
\bibliographystyle{frontiersinSCNS_ENG_HUMS} 
\bibliography{refs}


\end{document}